\documentclass[conference]{IEEEtran}
\IEEEoverridecommandlockouts

\usepackage{cite}
\usepackage{amsmath,amssymb,amsfonts}
\usepackage{algorithmic}
\usepackage{graphicx}
\usepackage{textcomp}
\usepackage{xcolor}
\def\BibTeX{{\rm B\kern-.05em{\sc i\kern-.025em b}\kern-.08em
    T\kern-.1667em\lower.7ex\hbox{E}\kern-.125emX}}
\begin{document}

\newcommand{\secref}[1]{Sec.~\ref{#1}}
\newcommand{\figref}[1]{Fig.~\ref{#1}}
\newcommand{\tabref}[1]{Table~\ref{#1}}
\newcommand{\eqnref}[1]{Eq.~\ref{#1}}

\title{Associating Healthcare Teamwork with Patient Outcomes for Predictive Analysis}

\author{\IEEEauthorblockN{Hsiao-Ying Lu}
\IEEEauthorblockA{\textit{Department of Computer Science} \\
\textit{University of California, Davis}\\
Davis, USA \\
hyllu@ucdavis.edu}
\and
\IEEEauthorblockN{Kwan-Liu Ma}
\IEEEauthorblockA{\textit{Department of Computer Science} \\
\textit{University of California, Davis}\\
Davis, USA \\
klma@ucdavis.edu}
}

\maketitle

\begin{abstract}
    Cancer treatment outcomes are influenced not only by clinical and demographic factors but also by the collaboration of healthcare teams. However, prior work has largely overlooked the potential role of human collaboration in shaping patient survival. This paper presents an applied AI approach to uncovering the impact of healthcare professionals’ (HCPs) collaboration---captured through electronic health record (EHR) systems---on cancer patient outcomes. We model EHR-mediated HCP interactions as networks and apply machine learning techniques to detect predictive signals of patient survival embedded in these collaborations. Our models are cross validated to ensure generalizability, and we explain the predictions by identifying key network traits associated with improved outcomes. Importantly, clinical experts and literature validate the relevance of the identified crucial collaboration traits, reinforcing their potential for real-world applications. This work contributes to a practical workflow for leveraging digital traces of collaboration and AI to assess and improve team-based healthcare. The approach is potentially transferable to other domains involving complex collaboration and offers actionable insights to support data-informed interventions in healthcare delivery.
\end{abstract}


\newcommand{\ProbDensFunc}{f}
\newcommand{\ProbDensFuncAll}{\ProbDensFunc_\mathrm{all}}
\newcommand{\ProbDensFuncClassZero}{\ProbDensFunc_\mathrm{0}}
\newcommand{\ProbDensFuncClassOne}{\ProbDensFunc_\mathrm{1}}

\newcommand{\nInsts}{n}
\newcommand{\nInstsClassZero}{\nInsts_\mathrm{0}}
\newcommand{\nInstsClassOne}{\nInsts_\mathrm{1}}

\newcommand{\DensRatioFunc}{g}
\newcommand{\DensRatioFuncClassZero}{\DensRatioFunc_\mathrm{0}}
\newcommand{\DensRatioFuncClassOne}{\DensRatioFunc_\mathrm{1}}

\newcommand{\Point}{\mathbf{p}}

\newcommand{\TransRate}{r}

\newcommand{\CaptionGPcomparison}{Comparison of collaborations with and without General Practitioner (GP) involvement. The top row shows two Breast Cancer patients with different outcomes, and the bottom row shows two Colorectal Cancer patients. Networks with GP involvement (left) exhibit higher connectivity and well-defined subteam clusters (highlighted in bright blue), indicating the GP’s role in bridging healthcare professionals and facilitating coordinated communication. In contrast, networks without GP (right) are more fragmented, with sparse connections and an absence of cohesive clusters, reflecting less integrated collaboration.}

\newcommand{\CaptionNetworkEx}{The illustration of the attributed bipartite collaboration network.}

\newcommand{\CaptionAblateEdge}{
    Projection of the bipartite collaboration network into an HCP-only network. On the left, a valid edge is formed when there exists a directed information flow between HCPs mediated by a note (e.g., $HCP_A$ writes $Note_B$ that is subsequently accessed by $HCP_C$). On the right, no edge is created because the HCPs do not share a sequential, mediated interaction through the same note, and thus no information flow can be inferred. The same projection rule applies to the note-only network.
}

\newcommand{\CaptionTimeframes}{
    The key milestones in the data collection and processing timeframe. We divide the collected data at the nine-month post-diagnosis mark into an observation window and a gap window. The observation window is used for training, while data from the gap window is excluded to prevent data leakage. Records in the gap window may encode late-stage clinical decisions or deterioration closely tied to survival outcomes, which would otherwise allow the model to access signals from late-stage treatment directly tied to the outcome.
}

\newcommand{\CaptionSage}{
    Our prediction model uses GraphSAGE layers to aggregate information from neighboring nodes and generate graph embeddings. A fully connected layer is then applied to these embeddings to predict patient survival.
}

\newcommand{\CaptionPredResults}{
    The prediction results using three different models on collaboration networks for each cancer type are presented. Bold text highlights the highest prediction accuracy, while red text marks the second-best performer. The results demonstrate that our \textit{collaboration-only} model consistently achieves the best patient survival prediction across all three cancer types. 
}

\newcommand{\CaptionCorrResults}{
    The correlation analyses examine four additional confounding factors that might be associated with patient survival. The results indicate that these factors exhibit little to no correlation with patient survival, as evidenced by correlation coefficients lower than $0.3$. Therefore, the likelihood of these factors being the primary drivers influencing patient survival is minimal.
}

\section{Introduction}

Cancer outcomes are influenced by a complex interplay of clinical, demographic, and organizational factors. While considerable research has focused on patient-level variables such as age, disease stage, and comorbidities \cite{sogaard2013impact, brandt2015age}, less attention has been paid to the role of human collaboration---particularly how healthcare professionals (HCPs) work together to coordinate and deliver care. Yet, effective collaboration is increasingly recognized as a critical determinant of care quality and patient outcomes in oncology and other complex medical domains \cite{gurses2006systematic, smits2010exploring, bagnasco2013identifying, verhaegh2017exploratory}.

Electronic Health Records (EHRs) play a central role in modern healthcare coordination, serving not only as repositories of patient information but also as platforms through which clinicians communicate, document, and implicitly collaborate \cite{blumenthal2010meaningful,RN249,RN244}. These digital traces of interaction offer a rich, but currently underutilized source of data for understanding how care teams function and how their collaboration may have associations with patient trajectories.

In this study, we propose a novel, data-driven workflow to evaluating team-based cancer care by modeling HCP collaboration patterns as networks derived from EHR interactions. We then apply machine learning techniques to these networks to identify predictive signals associated with patient survival. This approach allows us to move beyond static clinical indicators (i.e., cancer stage) and uncover human factors that associate with patient outcomes. By using cross-validation and explainable AI techniques, we ensure that our findings are both generalizable and actionable.

Feedback from medical experts indicated that our findings aligned with a long-standing hypothesis: involving general practitioners in coordinating cancer treatment can positively affect patient outcomes. A review of the literature \cite{smith2017shared, perfors2019involving, goderis2010start} confirmed that this hypothesis had been proposed but lacked empirical validation using real-world cancer treatment data. Our results provide the first data-driven evidence supporting it, reinforcing the practical relevance and credibility of our approach.

Our contributions are threefold: (1) we introduce a transferrable framework for extracting and modeling EHR-mediated collaboration networks; 
(2) we present an approach to analyzing medical collaboration data that mitigate spurious signals, and demonstrate how machine learning models can be applied to the resulting network representations to predict patient survival;
and (3) we provide evidence that the learned features offer actionable insights for improving care delivery. More broadly, this work showcases how leveraging digital traces of collaboration and AI can potentially support the assessment and optimization of team-based care in oncology, with potential applications across other high-stakes clinical domains.

\section{Related Works}
Cancer outcome prediction has traditionally focused on patient-level variables such as age, cancer stage, comorbidities, and treatment history \cite{piccirillo2004prognostic, sogaard2013impact, brandt2015age, nixon1994relationship}. These models have achieved varying levels of success in forecasting survival, but they often treat patients as isolated entities and overlook the broader context of care delivery and team-based coordination.

In parallel, there is growing recognition that collaboration among healthcare professionals can affect patient outcomes. Existing approaches include analyzing communication pathways and assessing the composition and roles of care teams \cite{gurses2006systematic, smits2010exploring, bagnasco2013identifying, verhaegh2017exploratory}. Despite this growing body of work, the role of team collaboration is still rarely integrated into predictive modeling pipelines.

Electronic health records (EHRs) offer a promising avenue for capturing these collaborative practices. Now pervasive in modern healthcare, EHRs serve as the primary platform for documenting, coordinating, and delivering care. Unlike surveys or observational studies, EHRs provide fine-grained, time-stamped logs that capture how healthcare professionals interact around patient care. While most research has focused on patient-level physiological data within EHRs \cite{huang2017regularized, amirahmadi2023deep, nelson2019integrating}, fewer studies have explored the metadata generated through system usage---such as access logs, shared documentation, or co-signatures---as signals of clinical collaboration. These underutilized digital traces can provide a nuanced and scalable view of how care teams function in real-world settings.

Machine learning has been increasingly applied to healthcare problems, including disease prediction, treatment recommendation, and patient risk stratification \cite{uddin2019comparing, atan2018deep, ballinger2018deepheart, rath2022prediction}. Some efforts have introduced explainability techniques to improve model transparency in clinical contexts \cite{mienye2024optimized, alsaleh2023prediction}. However, most of these works have focused on patient-level signals, with limited attention to modeling human factors such as collaboration among care providers. In our work, we apply machine learning to model EHR-mediated collaboration and address the unique challenges posed by this setting---such as the risk of spurious associations. Our approach emphasizes careful data modeling to ensure predictive validity and generalizability, offering a framework that can extend to other domains involving complex, team-based decision-making.

        \begin{figure}
            \centering
            \includegraphics[width=\linewidth]{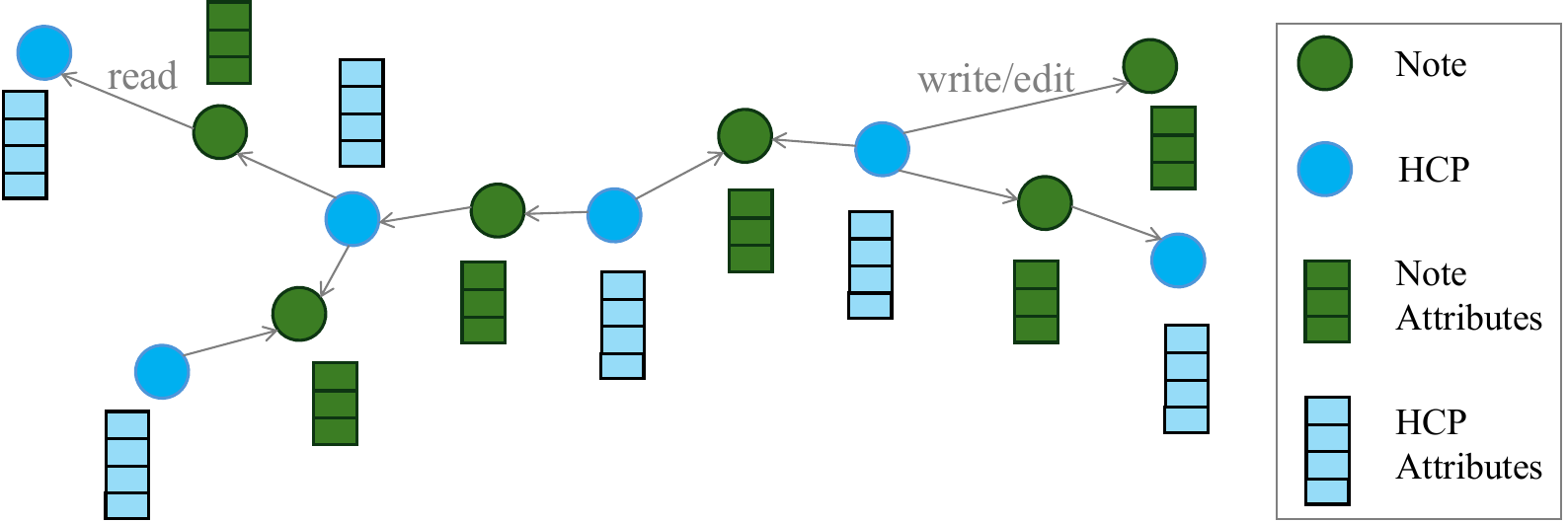}
            \caption{\CaptionNetworkEx{}}
            \label{fig:networkExample}
        \end{figure}
\section{Methodology}
    \subsection{Data: collaboration among HCPs through EHR}
        \textbf{Data overview.}
            The University of California Institutional Review Board approved our research (IRB \#1988738). Our raw data consists of EHR digital traces of patients diagnosed with Stage 2 or 3 breast, lung, and colorectal cancers. For each patient, the dataset includes their \textit{basic information} and \textit{access logs} of their EHR data. Basic information encompasses demographics (e.g., age and gender), treatments, comorbidities, and survival outcome (alive/dead). EHR access logs contain timestamped events spanning three months before to one year after the diagnosis date. 
            A timestamped EHR access event involves a healthcare professional (HCP) accessing (reviewing or writing) a document, such as a note or a message, on the EHR system. Since HCPs typically record a patient's medical conditions in notes, while messages often lack context, we include only the access events involving notes in our analysis. 
            To focus on assessing the interactions within core teams, following the recommendations of our medical doctor collaborators, we include only HCPs with the titles Medical doctor (MD), Nurse Practitioner (NP), Physician Assistant (PA), Registered Nurse (RN), Pharmacy Technician, Pharmacist, and Case Manager.

            This data captures the flow of information among HCPs, allowing us to track which HCP authored a note and who subsequently read it. Those who read the note may then write additional notes, further disseminating the acquired information. We extract these collaborative interactions that enable information transfer. To ensure a consistent collaboration timeframe, we exclude patients who passed away within a year of diagnosis, resulting in a total of 505 patients in this study. While this may introduce survivor bias by underrepresenting patients with rapidly progressing disease, it enables a more controlled comparison among patients with similar observation windows. In this setting, the predictive signals captured by the model are more likely to reflect meaningful differences in collaboration patterns rather than being driven primarily by acute clinical severity.
            
        \textbf{Cohort selection.} 
            The study cohort consists of 505 patients diagnosed with Stage II or III breast, lung, or colorectal cancers, selected based on the availability of complete EHR access logs and clinical records within the defined study period. Patients were not randomly sampled; rather, the cohort reflects individuals treated at a single academic medical center with sufficient longitudinal data for analysis. As a result, the cohort may be subject to selection bias, including institutional practice patterns and data availability constraints. These factors may limit the generalizability of the findings to other healthcare settings.
    
        \textbf{Data processing and categorization.}
        \label{sec:featureVar}
            To extract the collaboration surrounding a patient, we identify all notes related to the patient and the HCPs who have reviewed, written, or edited these notes using EHR access logs. Each note is further characterized by three variables: the category of the note's intent, the category of its content, and a label indicating whether it was created during an inpatient period. There are five intent categories, including \textit{Orders} and \textit{Patient Clinical Information}, and 32 content categories, such as \textit{Order Canceled} and \textit{Note Signed}.
            For each HCP, we provide context into their role in the collaboration using four variables: title, type, specialty, and a label indicating whether they are a resident. There are seven titles (e.g., \textit{MD} or \textit{RN}), 12 types (e.g., \textit{Physician Faculty} or \textit{Physician Fellow}), and 71 specialties (e.g., \textit{Cardiology} or \textit{Dermatology}).
            
        \begin{figure*}[h]
            \centering
            \includegraphics[width=0.9\linewidth]{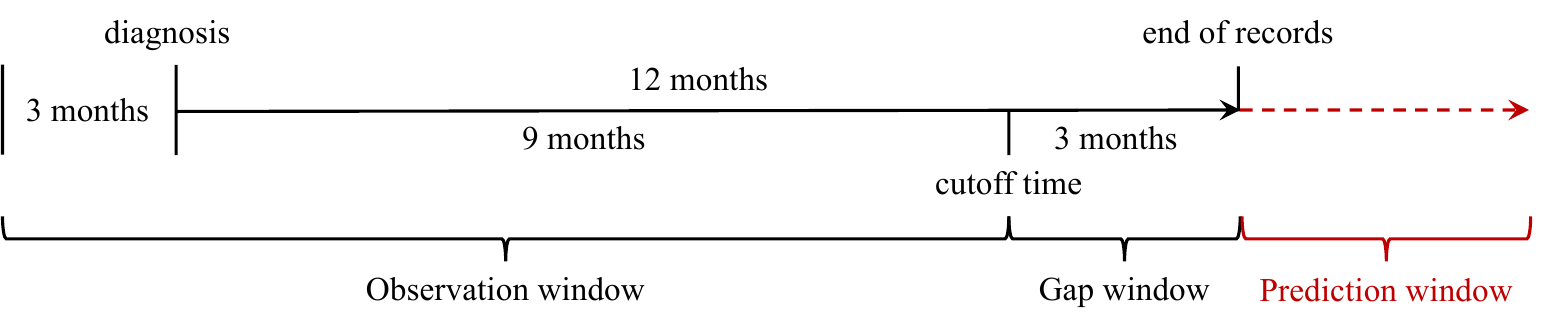}
            \caption{\CaptionTimeframes{}}
            \label{fig:timeframes}
        \end{figure*}
            
        \textbf{Bipartite network construction.}
            After identifying all participants in the collaboration surrounding a patient (notes and HCPs), we define the information flow among these entities. This flow is represented as a directed bipartite network, where notes and HCPs serve as network nodes, and edges capture the reviewing and writing events recorded in the EHR system. For example, if $HCP_A$ reviews $Note_B$, an edge is established from $Note_B$ to $HCP_A$, indicating that the information from $Note_B$ flows toward $HCP_A$.
            The bipartite nature of this collaboration network ensures that edges only form between a note and an HCP, but not between two notes or two HCPs. This reflects the fact that interactions among HCPs within the EHR are always mediated through notes rather than direct communication. Finally, the variables extracted to characterize each note and HCP are assigned as node attributes. The constructed collaboration networks are illustrated in \figref{fig:networkExample}. We adopt a stratified 5-fold cross-validation strategy, where patients are split into training and testing sets within each fold to ensure no overlap, helping the model learn generalizable characteristics rather than overfitting to previously seen data.


    \subsection{Prediction using Graph Neural Network}
        \textbf{Prediction variables and time windows.}
            Our goal is to explore whether, beyond the severity of a patient’s medical condition, how HCPs collaborate is associated with patient survival outcomes.
            Therefore, we train a machine learning model to predict a patient's survival outcome based on the traits of their collaboration network.

            We define three time windows to prevent data leakage in our predictive analysis. As shown in \figref{fig:timeframes}, the EHR log collection period spans from three months before to twelve months after a patient's diagnosis. We divide this period at the nine-month post-diagnosis mark into an observation window and a gap window. The EHR records from the observation window are used to train our prediction model, while those from the gap window are excluded.
            The decision to exclude the last three months of records aligns with the rationale for incorporating the three months preceding diagnosis, as this duration is considered significant in cancer treatment. Excluding the records from the gap window reduces the risk of information leakage, as later patterns of note access and editing may reflect care activities associated with a patient’s imminent survival status, which could confound the true patterns of collaboration we aim to study. Additionally, using only earlier data for prediction enables timely identification of underperforming collaborations, allowing for necessary interventions.
            It is important to note that all patients are still alive at the end of the recorded timeframe (i.e., twelve months post-diagnosis). The final survival outcome is determined at varying points after this period, depending on the patient. As a result, the length of the prediction window is not fixed due to the nature of this data. However, we ensure a consistent analysis timeframe for the observation and gap windows.
     
        \textbf{Model architecture.}
        \label{sec:predictionModel}
            \begin{figure*}
                \centering
                \includegraphics[width=\linewidth]{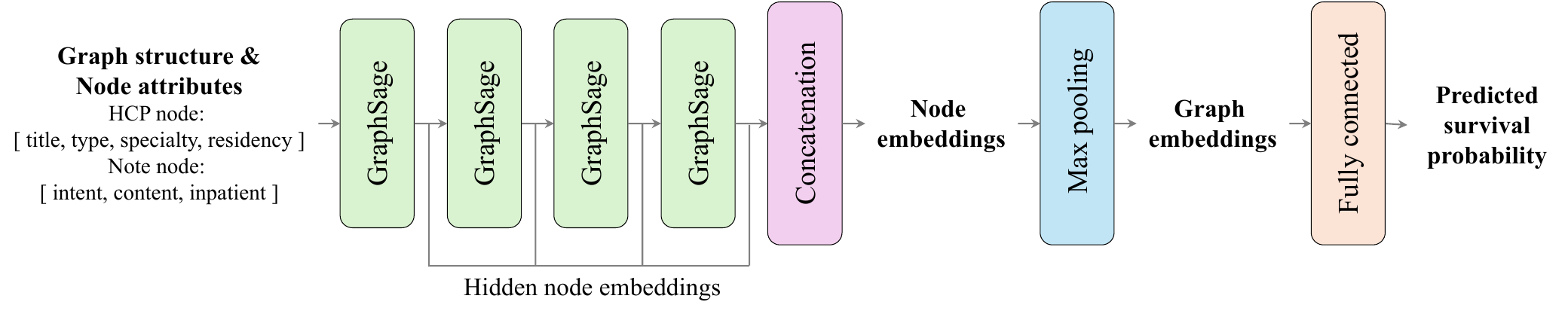}
                \caption{\CaptionSage{}}
                \label{fig:Sage}
            \end{figure*}
            Based on the defined time window configuration, we train a graph neural network (GNN) to learn from collaboration networks using only records from the observation window to predict patient survival outcomes. We use GraphSAGE \cite{hamilton2017inductive} to aggregate information from neighboring nodes in the collaboration network, leveraging their inductive capability to enhance the model’s generalizability to unseen patient collaborations in the future. As shown in \figref{fig:Sage}, the model architecture comprises four GraphSAGE layers followed by a fully connected prediction layer. The hidden outputs from each GraphSage layer are concatenated to form the final node embeddings, which are then max-pooled (i.e., aggregated by taking the maximum value of each dimension) across all nodes in the network to generate a graph embedding. This graph embedding is passed through the fully connected layer to produce the predicted probability of patient survival.

            Our GNN model captures the network traits shaped by both the collaboration topology and the node attributes, such as HCP specialty and note content, as detailed in \secref{sec:featureVar}. 
            The four-layer GraphSAGE architecture enables the model to capture collaboration patterns within four hops of information propagation, exemplified by paths such as \texttt{HCP-note-HCP-note-HCP} or \texttt{note-HCP-note-HCP-note}.

    \begin{figure}
        \centering
        \includegraphics[width=\linewidth]{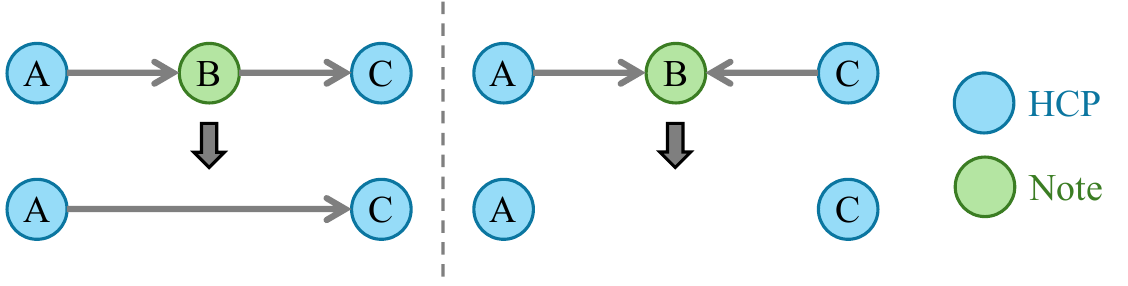}
        \caption{\CaptionAblateEdge{}}
        \label{fig:ablation}
    \end{figure}
    \subsection{Simplifications for explaining GNN predictions}
    \label{sec:ablatedModels}
        Our HCP collaboration networks pose unique challenges for explainability due to their complexity---particularly their bipartite and directed structure---on which existing explainable GNN methods \cite{ying2019gnnexplainer, luo2020parameterized, yuan2020xgnn} have not been directly evaluated or demonstrated consistent behavior. To address this, we disentangle and examine the two distinct sources of predictive signals separately: node attributes and network topology.

        \textbf{Node attributes.} To assess the potential association between node attributes and patient survival, we apply a max-pooling layer directly to the input attributes, bypassing the GraphSAGE and concatenation layers (see \figref{fig:Sage}). The pooled attributes are then fed into the fully connected prediction layer. This simplified architecture relies exclusively on node attributes for prediction, thereby eliminating interference from topological information.
        
        \textbf{Network topology.} To isolate predictive signals arising from topology, we simplify the collaboration networks to contain only HCPs or only notes, while removing all node attributes. In this case, predictions depend solely on structural relationships. For instance, as illustrated in \figref{fig:ablation}-left, a simplified edge is drawn from $HCP_A$ to $HCP_C$ if at least one note facilitates information flow between them. In this case, $Note_B$ mediates the information flow between $HCP_A$ and $HCP_C$. 
        The GNN is then trained on these reduced networks to learn exclusively from topological traits.
        
        By constraining each model to a single source of information---either attributes or topology---we approximate and explain the behavior of the full GNN. This separation allows us to attribute predictive power more precisely, enhancing interpretability.

        
        

    \subsection{Understanding predictions using explainable GNN methods}
    \label{sec:explainMethods}
        To help users interpret the GNN explanations, we leverage visual analytics techniques that enable interactive exploration of the model’s findings. For predictive signals originating from node attributes, we use NetworkCV \cite{lu2024visual}, which is specifically designed to explain neural network predictions on attribute-rich, multivariate networks, aligning with our simplified model architecture that focuses solely on node attributes. NetworkCV computes SHapley Additive exPlanations (SHAP) value \cite{lundberg2017unified} as a metric for quantifying the association between a node attribute and the patient survival.  
        To interpret the role of topological traits in the original collaboration networks, we employ GNNAnatomy \cite{lu2024gnnanatomy} to explain the GNN predictions made by using the simplified collaboration networks (i.e., excluding all node attributes). This tool allows us to pinpoint the topological signals contributing to the patient survival outcome predictions. 

        We explicitly disentangle the explanations for node attributes and for network topology to improve both interpretability and specificity. In contrast, most explainable GNN methods such as GNNExplainer \cite{ying2019gnnexplainer} use a generative model to jointly identify the most influential topology together with the attributes. However, this joint modeling does not quantify the relative importance of topology versus attributes, which can lead to ambiguous or inconsistent interpretations. Moreover, the influential subgraph identified by such methods reflects importance only within a particular graph instance, and its relevance may not generalize to other collaboration graphs.

            \begin{table*}
                \centering
                \caption{\CaptionPredResults{}}
                \includegraphics[width=\linewidth]{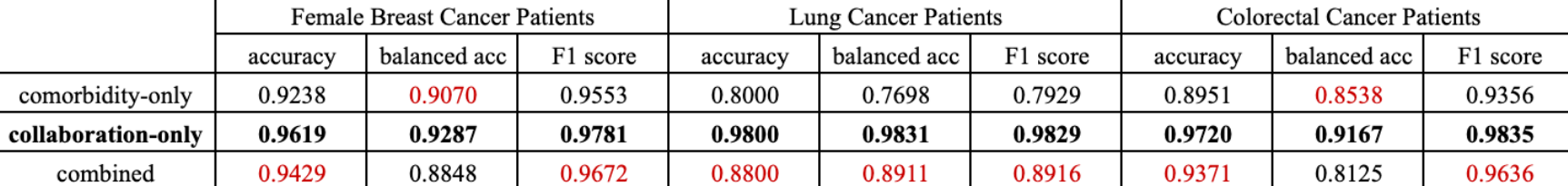}
                \label{tbl:predResults}
            \end{table*}
    \section{Experiments}
        \subsection{The importance of topological traits in collaborations among HCPs}
            Our goal is to explore whether the human factor---specifically, how HCPs collaborate---is associated with patient survival outcomes.
            However, other factors may implicitly influence the prediction. For instance, if a patient is more severely ill, HCPs may collaborate in a distinct manner. In this case, the true determinant of survival may not be the collaboration pattern itself but rather the patient’s level of sickness.
            To isolate the effect of HCP collaboration on patient survival and rule out as many confounding factors as possible, we use three different sets of data to conduct predictive analyses, supplemented by additional correlation analyses.

            \textbf{Predictive analyses.} 
            A common approach to assessing a patient’s general level of sickness is through comorbidities. Our comorbidity dataset includes 39 diseases, each represented by a binary label indicating whether the patient has been diagnosed with this condition. These 39 binary labels form a 39-dimensional comorbidity vector, which we use as a representation of patient sickness.
            
            We conduct three predictive analyses: one using only the comorbidity vector, one using only the collaboration, and one combining both. The model architecture described in \ref{sec:predictionModel} corresponds to the \textit{collaboration-only} approach. The \textit{comorbidity-only} model consists of a single fully connected layer followed by a nonlinear activation function. The \textit{combined} model follows the \textit{collaboration-only} architecture before the fully connected layer. The comorbidity vector is concatenated with the graph embedding before passed to the fully connected layer for final prediction.   
            
            Additionally, we separate the collaboration networks based on the patient's cancer type: breast cancer, lung cancer, and colorectal cancer. Since the teamwork patterns and the specialties of HCPs involved in treatment are expected to differ across cancer types, this separation ensures that the prediction model captures the unique collaboration patterns within each cancer type which influence patient survival. We also partition the data into separate training and testing sets for each cancer type, ensuring that the model learns generalizable characteristics rather than overfitting to previously seen data.

            \begin{table*}
                \centering
                \caption{\CaptionCorrResults{}}
                \includegraphics[width=\linewidth]{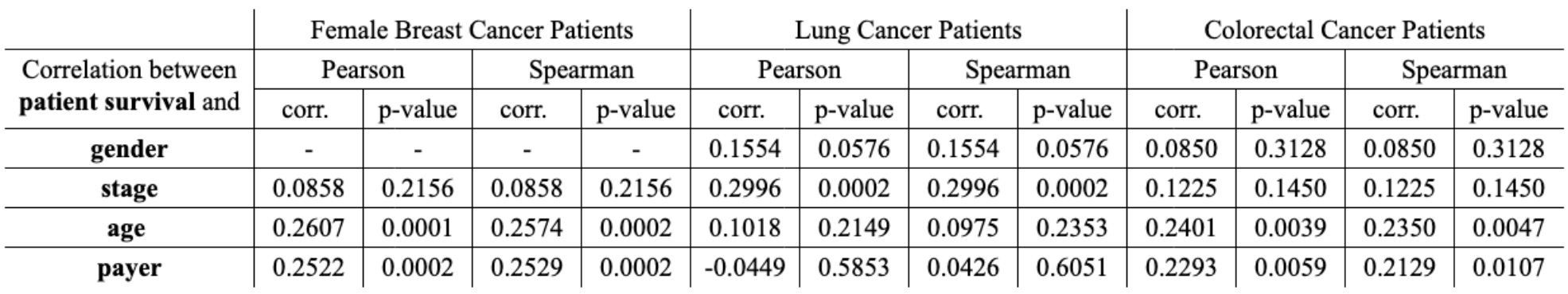}
                \label{tbl:corrResults}
            \end{table*}
            As shown in \tabref{tbl:predResults}, due to class imbalance, we report balanced accuracy and F1-score in addition to accuracy. Results are presented as mean ± standard deviation across 5 folds and despite the notably skewed class distribution, all models across different cancer cohorts achieve strong overall prediction performance.
            Notably, the \textit{collaboration-only} model outperforms all others across all three cancer datasets, followed by the \textit{combined} model, while the \textit{comorbidity-only} model demonstrates the lowest accuracy. These results suggest that while the comorbidity vector---representing a patient’s level of sickness---provides a reasonable indication of survival, it does not generalize well across all patients with the same cancer type. In contrast, the collaboration encodes more nuanced signals that contribute to more accurate predictions across different patients. 
            
            Interestingly, the \textit{combined} model does not outperform the \textit{collaboration-only} model, indicating that the signals extracted from comorbidity and collaboration are not well-aligned. This misalignment likely causes the \textit{combined} model to struggle in determining which signals to prioritize for prediction. Nevertheless, since the \textit{combined} model still outperforms the \textit{comorbidity-only} model, and the \textit{collaboration-only} model achieves the highest accuracy, 
            these findings highlight the strong predictive role of HCP collaboration in patient survival.

            \textbf{Correlation analyses.}
            To further ensure the effect of HCP collaboration on patient survival and minimize the influence of confounding factors, we examine four additional variables identified by our medical doctor collaborators as potentially correlated with patient survival: gender, cancer stage (Stage 2 or 3), age, and insurance type (private or public).
            We assess the alignment between each variable’s distribution and patient survival using both Spearman and Pearson correlation coefficients, along with their corresponding $p$-values. As shown in \tabref{tbl:corrResults}, the results from both correlation measures are highly consistent, providing nearly identical indications of the strength of association between these variables and patient survival. 
            
            Additionally, the highest correlation is observed between cancer stage and patient survival within the lung cancer dataset. However, even this highest correlation remains below $0.3$, indicating little to no meaningful association between the two variables. Furthermore, the relatively high $p$-values (some exceeding $0.6$) suggest that any weak correlation detected is likely due to chance rather than a true relationship.
            Thus, the likelihood of these four variables being primary drivers of patient survival is minimal. 
            Based on this, we find evidence that the association between HCP collaboration and patient survival is not readily explained by these examined factors.

            \begin{figure*}[h]
                \centering
                \includegraphics[width=0.78\linewidth]{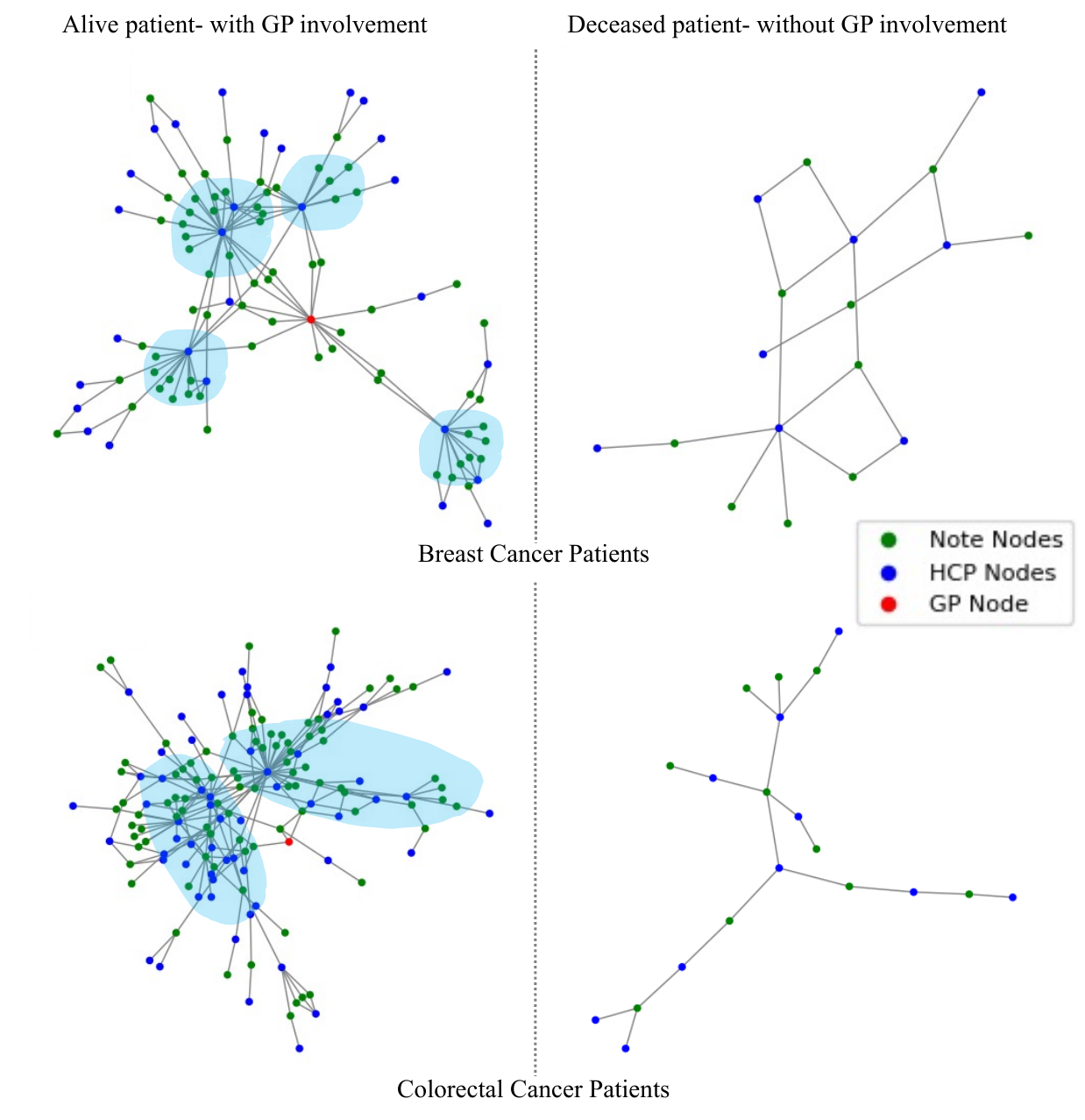}
                \caption{\CaptionGPcomparison{}}
                \label{fig:GPcomparison}
            \end{figure*}
            
        \subsection{The collaboration patterns affecting patient survival outcome }
            To understand the predictive signals captured by our GNN model, we use GNNAnatomy to explain the topological traits and NetworkCV to identify contributing node attributes, as introduced in \secref{sec:explainMethods}. GNNAnatomy \cite{lu2024gnnanatomy}, found that there are no dominant topology underlying the information flows among healthcare providers and clinical notes that directly contribute to patient survival. In contrast, NetworkCV \cite{lu2024visual} highlighted several non-trivial node attributes that point to meaningful collaboration patterns.
            
            In lung cancer, we observed that some NetworkCV-identified contributing HCP attributes are subtle indicators of comorbidities---such as the involvement of cardiologists or other specialists. Additionally, some important Note attributes suggest that patients whose providers infrequently accessed clinical notes tended to have poorer outcomes. Importantly, the range of SHAP values for these node attributes are comparable, implying that the model is not overly reliant on one trait to make predictions, but rather incorporates a diverse set of behavioral signals.

            For breast and colorectal cancers, the most influential HCP attribute is the involvement of general practitioners (GPs). The presence of a GP in the care team was consistently associated with improved patient outcomes. Notably, the SHAP value for GP involvement exceeded those of the subtle comorbidity signals---such as involvement of emergency medicine or cardiology. This highlights a particularly strong impact of GP participation. Additionally, this pattern holds across multiple cancer types, further suggesting that GP involvement may reflect a more holistic, coordinated care process that benefits patient survival. These findings offer evidence reinforcing the clinical hypothesis that generalist physicians play a critical role in cancer care.
            
            As shown in \figref{fig:GPcomparison}, comparison of patients with different survival outcomes reveals that the GP plays a central coordinating role across major collaboration clusters for patients who survived. In networks with GP involvement, the main collaboration clusters connected by the GP are highlighted in bright blue, indicating cohesive and well-integrated subteams. In contrast, patients without GP involvement exhibit few meaningful clusters, with collaboration largely confined to a small subset of healthcare providers (HCPs). This limited network connectivity suggests reduced coordination of care and restricted engagement with potentially relevant providers, which may be associated with the absence of GP involvement.

    
    \subsection{Evaluations and broader impact}
        In this study, we sought to isolate collaboration patterns associated with cancer survival while minimizing confounding factors (\tabref{tbl:corrResults}). To test robustness, we performed five-fold cross-validation, reporting average metrics across folds (\tabref{tbl:predResults}). This design reduces overfitting and shows that the learned predictive signals generalize to unseen patients, with consistent performance supporting the model’s stability.

        To assess the clinical relevance of our findings, we presented key model insights to domain experts in oncology and primary care. Experts confirmed that the involvement of general practitioners (GPs) in cancer care aligns with their hypothesis that GPs play a central role in coordinating treatment plans across multiple providers. This feedback echoes the broader hypothesis of shared care, where collaboration between primary care and specialist providers is known to improve the management of long-term and complex conditions. While this model of care has been advocated in the literature \cite{smith2017shared, perfors2019involving, goderis2010start, smith2008chronic, scherpbier2013effect, holm2002randomized, dey2002randomized, byng2004exploratory}, especially for chronic diseases, its application in cancer care has not been validated using real-world data. Our study offers the first empirical support for the effectiveness of shared care in oncology, providing evidence that involving general practitioners should be more formally considered in efforts to improve cancer treatment delivery and patient outcomes.
\section{Discussion}
    \textbf{Validation in Real Practice. }
        While our findings align with previous studies and expert feedback, they require validation in clinical settings. In particular, the association between collaboration patterns---such as GP involvement---and patient outcomes should be assessed by medical professionals before guiding interventions. Future work includes implementing trials to evaluate their causal impact on patient care and survival.

    \textbf{Time Window Length Exploration. }
        There is no standard for defining observation and prediction windows in HCP collaboration studies. Our chosen windows were guided by data availability and cancer treatment practices, making them effective for this context but not necessarily generalizable. They should be seen as empirically motivated design choices, not benchmarks. Future work should explore how varying window configurations affect the stability and interpretability of collaboration patterns.

    \textbf{Alternative Predictive Models. }
        Simple statistical models (e.g., logistic regression using individual features such as GP involvement) may achieve comparable predictive performance, suggesting that some signals associated with patient outcomes can be captured using relatively low-dimensional representations. This highlights that certain aspects of collaboration, such as the presence of specific roles, may already carry informative signals.
        At the same time, our graph-based approach enables the modeling of higher-order interaction structures among healthcare professionals, offering a complementary perspective on how collaboration unfolds in practice. These structures are difficult to represent using traditional feature-based models and may reflect more nuanced patterns of coordination.
        It is important to note that both statistical and graph-based approaches rely on observational data and are therefore subject to similar confounding limitations. As such, the results should be interpreted as identifying meaningful associations between collaboration patterns and patient outcomes, rather than establishing causal relationships.

\section{Conclusion}
    This paper presents a data-driven framework for modeling and analyzing healthcare professional collaboration using electronic health record (EHR) data to predict cancer patient survival. Our approach demonstrates robust and generalizable predictive performance through careful experimental design and evaluation, while also providing clinically meaningful insights supported by expert feedback and the healthcare literature. Notably, our findings offer the first empirical evidence for a long-standing hypothesis: the involvement of general practitioners plays a beneficial role in cancer care coordination and patient survival outcomes. By leveraging real-world data and machine learning, our work contributes a methodology for identifying actionable intervention targeting human factors in complex care collaborations, with potential applications beyond oncology.

\section*{Acknowledgment}
This work was supported by grant R01CA273058 and R01CA270454 from the National Cancer Institute. Contents of this manuscript are solely the responsibility of the authors and do not represent the official view of the National Cancer Institute.

\bibliographystyle{IEEEtran}
\bibliography{00_ref}

\end{document}